%% file: Asilomar2024_arxiv_V2.tex
\documentclass[a4paper,twocolumn, 10pt]{IEEEtran} 

\usepackage{amsmath}
\usepackage{amssymb}
\usepackage{graphicx}
\usepackage{color} 
\usepackage{xcolor}
\usepackage{cite}
\usepackage{bm}
\usepackage{tabularx}
\usepackage{acronym}
\usepackage[yyyymmdd,hhmmss]{datetime}
\usepackage{bbm}
\usepackage{notation}
\usepackage{mathrsfs} 
\usepackage{bardiag}
\usepackage{tabularx}
\usepackage{multirow}
\usepackage{colortbl,pgfplotstable}
\usepackage{mathtools, cuted}
\usepackage{xr}
\usepackage{pdfpages}

\usepackage{dblfloatfix}

\usepackage{multirow}
\usepackage{makecell}
\usepackage{booktabs}

\usepackage{enumitem} 
\usepackage[T1]{fontenc}
\usepackage[utf8]{inputenc}
\usepackage{pgfplots}
\usepackage{grffile}
\pgfplotsset{compat=newest}
\usetikzlibrary{plotmarks}
\usetikzlibrary{arrows.meta}
\usepgfplotslibrary{patchplots}

\usepackage[caption=false,font=footnotesize]{subfig}
\usepackage{pgfplots}
\usepackage{tikzscale}
\newcommand{\exportFigures}{true} %
\newcommand{\exportFiguresAsPNG}{false}

\newcommand{\tikzfolder}{./compiledPlots/}

\input{pgfplot}

\allowdisplaybreaks
\newcolumntype{L}[1]{>{\raggedright\arraybackslash}p{#1}}
\newcolumntype{C}[1]{>{\centering\arraybackslash}p{#1}}
\newcolumntype{R}[1]{>{\raggedleft\arraybackslash}p{#1}}

\providecommand{\ist}{\hspace*{.3mm}}

\providecommand{\rmv}{\hspace*{-.3mm}}
\providecommand{\iist}{\hspace*{1mm}}

\providecommand{\nn}{\nonumber}

\newcommand{\T}{\text{T}}

\newcommand{\atantwo}{\text{atan2}}

\newtheorem{problem*}{Problem}

\newdateformat{monthyeardate}{\monthname[\THEMONTH] \THEDAY, \THEYEAR} %

\definecolor{colA}{rgb}{0,0,1}%
\definecolor{colB}{rgb}{0.8,0.0,0.5}%
\definecolor{colC}{rgb}{0,0.5,0}%
\definecolor{col_cyan}{rgb}{0,1,1}%
\definecolor{colE}{rgb}{1.00000,0.55,0.00000}%
\definecolor{colF}{rgb}{1.00000,0.0,0.00000}%
\definecolor{col_mag}{rgb}{1.00000,0.00000,1.00000}%
\definecolor{col_grey}{rgb}{0.4,0.4,0.4}%
\definecolor{col_gray2}{rgb}{0.4,0.4,0.4}
\definecolor{lightgray}{rgb}{0.9,0.9,0.9}

\definecolor{col_grey2}{rgb}{0.6,0.6,0.6}
\definecolor{Gray}{rgb}{225, 225, 225} 
\definecolor{matlabBlue}{rgb}{0.00000,0.44700,0.74100}%
\definecolor{matlabOrange}{rgb}{0.85000,0.32500,0.09800}%
\definecolor{matlabYellow}{rgb}{0.92900,0.69400,0.12500}%
\definecolor{matlabLila}{rgb}{0.49400,0.18400,0.55600}%
\definecolor{matlabGreen}{rgb}{0.46600,0.67400,0.18800}%
\definecolor{col_PAgreen}{rgb}{0.00000,0.49800,0.00000}%

\definecolor{alex}{RGB}{235,134,52}
\definecolor{erik}{RGB}{235,134,52}

\definecolor{FGgreen}{RGB}{34,139,34}
\definecolor{FGblue}{RGB}{80,120,255}
\definecolor{FGred}{RGB}{255,110,110}

\colorlet{fgColor_node}{colA}
\colorlet{fgColorUp1}{colB}
\colorlet{fgColor_coop}{colB}
\colorlet{fgColorBG}{colC}
\colorlet{fgColor_facAlpha}{colE}
\colorlet{fgColorPost3}{colF}
\colorlet{fgColorComb}{colE}
\colorlet{fgColorBox}{colE}

\def\linewidthA{1}

\def\mylinewidth{0.7}
\def\mylinewidth2{1}

\def\figureW{6cm}
\def\figureW2{5cm}

\def\figureDataW{5cm}
\def\figureDataH{4cm}

\def\marksizeA{1.5}

\pgfplotsset{styleE1/.style={mymarkfixednumber={\markD}{mark size=\marksizeA,solid}{10},color=matlabGreen,densely dashed,  line width=\linewidthA, mark repeat=1, mark phase = 0, mark options={solid,matlabGreen}}}

\pgfplotsset{styleE2/.style={mymarkfixednumber={\markDd}{mark size=\marksizeA,solid}{10},color=matlabGreen, line width=\linewidthA, mark repeat=1, mark phase = 0, mark options={solid,matlabGreen}}}

\pgfplotsset{styleE3/.style={mymarkfixednumber={\markBb}{mark size=\marksizeA,solid}{10},color=matlabLila, line width=\linewidthA, mark repeat=1, mark phase = 0, mark options={solid,matlabLila}}}

\pgfplotsset{styleE4/.style={mymarkfixednumber={\markC}{mark size=\marksizeA,solid}{10},color=matlabOrange,densely dotted, line width=\linewidthA, mark repeat=1, mark phase = 0, mark options={solid,matlabOrange}}}

\pgfplotsset{styleE5/.style={mymarkfixednumber={\markCc}{mark size=\marksizeA,solid}{10},color=matlabOrange, line width=\linewidthA, mark repeat=1, mark phase = 0, mark options={solid,matlabOrange}}}

\pgfplotsset{styleE6/.style={mymarkfixednumber={\markA}{mark size=\marksizeA,solid}{10},color=matlabBlue,densely dotted, line width=\linewidthA, mark repeat=1, mark phase = 0, mark options={solid,matlabBlue}}}

 \pgfplotsset{styleE7/.style={mymarkfixednumber={\markAa}{mark size=\marksizeA,solid}{12},color=matlabBlue, line width=\linewidthA, mark repeat=1, mark phase = 0, mark options={solid,matlabBlue}}}

\pgfplotsset{styleE8/.style={mymarkfixednumber={\markA}{mark size=\marksizeA,solid}{6},color=matlabBlue,densely dashed, line width=\linewidthA, mark repeat=1, mark phase = 0, mark options={solid,matlabBlue}}}

\newcommand{\zd}{\ensuremath{z^{(j,i)}_{\text{d}_{m,n}}}}
\newcommand{\zu}{\ensuremath{z^{(j,i)}_{\mathrm{u}_{m,n}}}}

\newcommand{\zaoa}{\ensuremath{z^{(j,i)}_{\theta_{m,n}}}}
\newcommand{\zaod}{\ensuremath{z^{(j,i)}_{\vartheta_{m,n}}}}

\graphicspath{{./plots/}}

\input{\folder/acronyms.tex}

\IEEEoverridecommandlockouts

\begin{document}
\frenchspacing

\title{\LARGE 
Multipath-based SLAM with Cooperation and Map Fusion\\in MIMO Systems}
\author{
Erik Leitinger, Lukas Wielandner,
Alexander Venus,
and Klaus Witrisal\\[2mm] 
Graz University of Technology
}
\maketitle

\begin{abstract}
  \input{\folder/abstract.tex}
\end{abstract}  
                 

\acresetall

\vspace*{-3mm}

\section{Introduction}\label{sec:introduction}

\input{\folder/introduction.tex}
\input{\folder/sigmod.tex}

\input{\folder/sysmodslam.tex}

\input{\folder/scheduling_and_MP.tex}
\input{\folder/simulation_results.tex}

\vspace*{-2mm}
\section{Conclusions}\label{sec:conclusion}
\input{\folder/conclusions.tex}

\appendices
\input{\folder/Appendix.tex}


\bibliographystyle{IEEEtran}
\bibliography{IEEEabrv,references}
\end{document}

%% file: pgfplot.tex
\usetikzlibrary{shapes}
\usetikzlibrary{shapes.misc}
\usetikzlibrary{arrows}
\usetikzlibrary{arrows.meta}
\usetikzlibrary{angles}
\usetikzlibrary{quotes}
\usetikzlibrary{fit}
\usetikzlibrary{graphs}
\usetikzlibrary{patterns}
\usetikzlibrary{petri}
\usetikzlibrary{plotmarks}
\usetikzlibrary{positioning}
\usetikzlibrary{decorations.pathmorphing}
\usetikzlibrary{decorations.markings}
\usetikzlibrary{decorations.pathreplacing}
\usetikzlibrary{decorations.shapes}
\usetikzlibrary{decorations.text}
\usetikzlibrary{backgrounds}
\usetikzlibrary{calc}
\usetikzlibrary{spy}
\usetikzlibrary{math}

\usepgfplotslibrary{patchplots}
\usepgfplotslibrary{colormaps}

\pgfdeclarelayer{backback1}
\pgfdeclarelayer{back1}
\pgfdeclarelayer{lay1}
\pgfdeclarelayer{backback2}
\pgfdeclarelayer{back2}
\pgfdeclarelayer{lay2}
\pgfdeclarelayer{frontlay}
\pgfsetlayers{backback2,back2,lay2,backback1,back1,lay1,main, frontlay}

\ifthenelse{\equal{\exportFigures}{true}}
{
	\usepgfplotslibrary{external}
	\tikzexternalize[prefix=\tikzfolder,optimize command away=\includepdf]
	\ifthenelse{\equal{\exportFiguresAsPNG}{true}}
	{
		\tikzset
		{   png export/.style={
				external/system call={
					pdflatex \tikzexternalcheckshellescape 
					-halt-on-error 
					--extra-mem-top=10000000 
					-interaction=batchmode 
					-jobname "\image" "\texsource" && pdftops -eps "\image.pdf" && convert -density 700 -transparent white "\image.pdf" "\image.png"
		}}}
		\tikzset{png export}
	}
	{}
}
{}

\def\addlegendimage{\csname pgfplots@addlegendimage\endcsname}
\def\addlegendentry{\csname pgfplots@addlegendentry\endcsname}

\newcommand{\pgfref}[1]
{\ifthenelse{\equal{\exportFigures}{true}}
{\tikzexternaldisable\ref{#1}\tikzexternalenable}
{\ref{#1}}}

\pgfplotsset{compat=newest} 

\usetikzlibrary{backgrounds}
\tikzstyle{load}   = [ultra thick,-latex]
\tikzstyle{stress} = [-latex]
\tikzstyle{dim}    = [latex-latex]
\tikzstyle{axis}   = [-latex,black!55]

\definecolor{applegreen}{rgb}{0.55, 0.71, 0.0}
\definecolor{asparagus}{rgb}{0.53, 0.66, 0.42}
\definecolor{darkspringgreen}{rgb}{0.09, 0.45, 0.27}
\definecolor{emerald}{rgb}{0.31, 0.78, 0.47}
\definecolor{forestgreen}{rgb}{0.13, 0.55, 0.13}
\definecolor{green(ryb)}{rgb}{0.4, 0.69, 0.2}
\definecolor{green(pigment)}{rgb}{0.0, 0.65, 0.31}

\definecolor{mediumpurple}{rgb}{0.58, 0.44, 0.86}
\definecolor{mediumred-violet}{rgb}{0.73, 0.2, 0.52}

\definecolor{frenchblue}{rgb}{0.0, 0.45, 0.73} 

\definecolor{mediumcandyapplered}{rgb}{0.89, 0.02, 0.17}

\definecolor{pumpkin}{rgb}{1.0, 0.46, 0.09}

\usetikzlibrary{decorations.pathreplacing,decorations.markings}
\usetikzlibrary{arrows.meta}

\newlength\figureheight 
\newlength\figurewidth 
\usetikzlibrary{patterns,arrows}
\usepgfplotslibrary{colormaps}
\usetikzlibrary{plotmarks}
\pgfplotsset{every axis/.append style={
  label style={font=\normalsize},
  tick label style={font=\scriptsize},
  xticklabel={
   \ifdim \tick pt < 0pt
    \pgfmathparse{abs(\tick)}%
    \llap{$-{}$}\pgfmathprintnumber{\pgfmathresult}
   \else
    \pgfmathprintnumber{\tick}
   \fi}
   }}
\usetikzlibrary{decorations.markings}
\makeatletter
\tikzset{
  nomorepostactions/.code={\let\tikz@postactions=\pgfutil@empty},
  mymarkfixednumber/.style n args={3}{decoration={markings,
    mark= between positions 0.1 and 1 step (1/#3)*\pgfdecoratedpathlength with{%
        \tikzset{#2,every mark}\tikz@options
        \pgfuseplotmark{#1}%
      },  
    },
    postaction={decorate},
    /pgfplots/legend image post style={
        mark=#1,mark options={#2},every path/.append style={nomorepostactions}
    },
  },
}
\tikzset{
  nomorepostactions/.code={\let\tikz@postactions=\pgfutil@empty},
  mymarkfixeddistance/.style n args={3}{decoration={markings,
    mark= between positions 0 and 1 step #3cm with{%
        \tikzset{#2,every mark}\tikz@options
        \pgftransformresetnontranslations%
        \pgfuseplotmark{#1}%
      },  
    },
    postaction={decorate},
    /pgfplots/legend image post style={
        mark=#1,mark options={#2},every path/.append style={nomorepostactions}
    },
  },
}
\tikzset{
  nomorepostactions/.code={\let\tikz@postactions=\pgfutil@empty},
  mymark/.style n args={3}{decoration={markings,
    mark= between positions 0 and 1 step 0.75cm with{%
        \tikzset{#2,every mark}\tikz@options
        \pgftransformresetnontranslations%
        \pgfuseplotmark{#1}%
      },  
    },
    postaction={decorate},
    /pgfplots/legend image post style={
        mark=#1,mark options={#2},every path/.append style={nomorepostactions}
    },
  },
}
\makeatother

\definecolor{colA}{rgb}{0,0,1}%
\definecolor{colB}{rgb}{1.00000,0.0,0.00000}%
\definecolor{colC}{rgb}{0,0.7,0}%
\definecolor{colD}{rgb}{0.9,0.0,0.9}%
\definecolor{colE}{rgb}{0,1,1}%
\definecolor{colF}{rgb}{1.00000,0.55,0.00000}%

\def\mylinewidth{0.4}

\def\markA{o}
\def\markAa{*}

\def\markBb{square*}
\def\markC{triangle}
\def\markCc{triangle*}
\def\markD{diamond}
\def\markDd{diamond*}

\pgfdeclareplotmark{markarrow}
{%
  \pgfpathmoveto{\pgfqpoint{\pgfplotmarksize}{0pt}}
  \pgfpathlineto{\pgfqpointpolar{130}{1.5\pgfplotmarksize}}
  \pgfpathmoveto{\pgfqpoint{\pgfplotmarksize}{0pt}}
  \pgfpathlineto{\pgfqpointpolar{-130}{1.5\pgfplotmarksize}}
  \pgfusepathqstroke
}

%% file: inputFiles/acronyms.tex
\acrodefplural{esl}[ESLs]{electronic shelf labels}

\acrodef{rf}[RF]{radio frequency}
\acrodef{bs}[BS]{base station}
\acrodef{mt}[MT]{mobile terminal}
\acrodef{aoa}[AOA]{angle-of-arrival}
\acrodef{aod}[AOD]{angle-of-departure}
\acrodef{awgn}[AWGN]{additive white Gaussian noise}
\acrodef{blt}[BLT]{bluetooth}
\acrodef{bp}[BP]{belief propagation}
\acrodef{bw}[BW]{bandwidth}
\acrodef{cdf}[CDF]{cumulative distribution function}
\acrodef{ceda}[CEDA]{channel estimation and detection algorithm}
\acrodef{cf}[CF]{cumulative frequency}
\acrodef{ci}[CI]{confidence interval}
\acrodef{cl}[CL]{confidence level}
\acrodef{crlb}[CRLB]{Cram\'er-Rao lower bound}
\acrodef{dmc}[DMC]{dense multipath component}
\acrodef{dm}[DM]{dense multipath}
\acrodef{dnr}[DNR]{dense-to-noise ratio}
\acrodef{dps}[DPS]{delay power spectrum}
\acrodef{dut}[DUT]{device under test}
\acrodef{eirp}[EIRP]{equivalent isotropic radiated power}
\acrodef{eot}[EOT]{extended object tracking}
\acrodef{fa}[FA]{false alarm}
\acrodef{Fa}[FA]{False alarm}
\acrodef{fg}[FG]{factor graph}
\acrodef{glrt}[GLRT]{generalized likelihood ratio test}
\acrodef{iid}[iid]{independent and identically distributed}
\acrodef{imu}[IMU]{inertial measurement unit}
\acrodef{lhf}[LHF]{likelihood function}
\acrodef{los}[LOS]{line-of-sight}
\acrodef{lpbo}[LPBO]{legacy \ac{pbo}}
\acrodef{mc}[MC]{Monte Carlo}
\acrodef{mf}[MF]{matched filter}
\acrodef{mie}[MIE]{method of interval estimation}
\acrodef{mimo}[MIMO]{multiple input multiple output}
\acrodef{ml}[ML]{maximum likelihood}
\acrodef{mmse}[MMSE]{minimum mean-square error}
\acrodef{mpc}[MPC]{multipath component}
\acrodef{mp}[MP]{multipath}
\acrodef{MP}[MP]{message passing}
\acrodef{mpslam}[MP-SLAM]{multipath-based simultaneous localization and mapping}
\acrodef{Mpslam}[MP-SLAM]{Multipath-based simultaneous localization and mapping}
\acrodef{mse}[MSE]{mean squared error}
\acrodef{mse}[MSE]{mean squared error}
\acrodef{mva}[MVA]{master \ac{va}}
\acrodef{nlike}[NLIKE]{normalized likelihood}
\acrodef{nlos}[NLOS]{non-\ac{los}}
\acrodef{nnlike}[NNLIKE]{normalized noise-free likelihood}
\acrodef{npbo}[NPBO]{new \ac{pbo}}
\acrodef{olos}[OLOS]{obstructed \ac{los}}
\acrodef{ospa}[OSPA]{optimal subpattern assignment}
\acrodef{mospa}[MOSPA]{mean \ac{ospa}}
\acrodef{pa}[PA]{physical anchor}
\acrodef{pbo}[PBO]{potential bias object}
\acrodef{pcb}[PCB]{printed circuit board}
\acrodef{pcrlb}[PCRLB]{posterior Cram\'er-Rao lower bound}
\acrodef{pdaai}[AIPDA]{amplitude-information \ac{pda}}
\acrodef{pdaf}[PDAF]{probabilistic data association filter}
\acrodef{pda}[PDA]{probabilistic data association}
\acrodef{pdf}[PDF]{probability density function}
\acrodef{pdp}[PDP]{power delay profile}
\acrodef{pmf}[PMF]{probability mass function}
\acrodef{pva}[PVA]{potential \ac{va}}
\acrodef{reb}[REB]{ranging error bound}
\acrodef{rmse}[RMSE]{root mean squared error}
\acrodef{rss}[RSS]{received signal strength}
\acrodef{rv}[RV]{random variable}
\acrodef{simo}[SIMO]{single input multiple output}
\acrodef{sinr}[SINR]{signal-to-interference-plus-noise-ratio}
\acrodef{slam}[SLAM]{simultaneous localization and mapping}
\acrodef{smc}[SMC]{specular multipath component}
\acrodef{snr}[SNR]{signal-to-noise-ratio}
\acrodef{spa}[SPA]{sum-product algorithm}
\acrodef{stdv}[STDV]{standard deviation}
\acrodef{tdoa}[TDOA]{time difference of arrival}
\acrodef{tka}[TKA]{trusted keyless access}
\acrodef{toa}[TOA]{time-of-arrival}
\acrodef{ula}[ULA]{uniform linear array}
\acrodef{ura}[URA]{uniform rectangular array}
\acrodef{ut}[UT]{unscented transform}
\acrodef{uwb}[UWB]{ultra wide band}
\acrodef{va}[VA]{virtual anchor}
\acrodef{zzb}[ZZB]{Ziv-Zakai bound}

%% file: inputFiles/abstract.tex
\ac{Mpslam} is a promising approach in wireless networks for obtaining position information of transmitters and receivers as well as information on the propagation environment. \ac{Mpslam} models specular reflections of \ac{rf} signals at flat surfaces as \acfp{va}, the mirror images of \acp{bs}. Conventional methods for \ac{Mpslam} consider a single \ac{mt} which has to be localized. The availability of additional \acp{mt} paves the way for utilizing additional information in the scenario. Specifically enabling \acp{mt} to exchange information allows for data fusion over different observations of \acp{va} made by different MTs. Furthermore, cooperative localization becomes possible in addition to multipath-based localization. Utilizing this additional information enables more robust mapping and higher localization accuracy.

%% file: inputFiles/introduction.tex
\begin{figure}[!t]
	\centering   
	\includegraphics{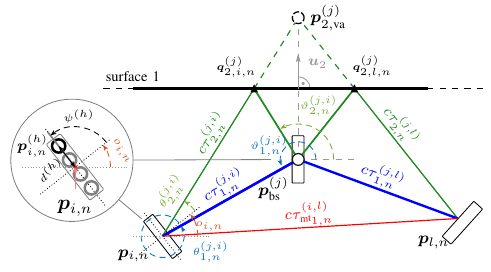}
	\caption{Exemplary indoor environment including a \ac{bs} at position \smash{$\V{p}^{(j)}_\text{bs} $} and two MTs at position $\V{p}_{k,n}$, sharing information about a \ac{va} at position \smash{$\V{p}^{(j)}_{l,\text{va}}$}. A visualization of the array geometry definition used at the agent and \acp{pa} is included.}
	\vspace*{-4mm}
	\label{fig:scenario}
\end{figure}

\ac{Mpslam} is a powerful approach in wireless networks for obtaining position information of transmitters and receivers as well as information on the propagation environment. \ac{Mpslam} models specular reflections of \ac{rf} signals at flat surfaces as \acfp{va}, the mirror images of \acp{bs} as shown in Fig.~\ref{fig:scenario} \cite{WitMeiLeiSheGusTufHanDarMolConWin:J16}. \ac{Mpslam} can detect and localize \acp{va} and jointly estimate the time-varying \ac{mt} position \cite{WitMeiLeiSheGusTufHanDarMolConWin:J16, GentnerTWC2016, LeiMeyHlaWitTufWin:J19, LeiGreWit:ICC2019,MenMeyBauWin:J19}. The availability of \ac{va} location information makes it possible to leverage multiple propagation paths of \ac{rf} signals for \ac{mt} localization, thus significantly improving localization accuracy and robustness \cite{LeitingerJSAC2015,WilGreLeiMueWit:ACSSC2018, ShaGarDesSecWym:TWC2018, MenWymBauAbu:TWC2019}.

\subsection{State of the Art}
\ac{Mpslam} is a feature-based SLAM approach that focuses on detecting and mapping distinct features in the environment \cite{DurrantWhyte2006,MonThrKolWeg:AAAI2002,DeuReuDie:SPL2015}. Features of interest are often referred to as \acp{va}\cite{LeiMeyHlaWitTufWin:J19,LeiGreWit:ICC2019,MenMeyBauWin:J19, ChuLuGesWanWenWuMuqLi:TWC2022} and `measurements'' are obtained by extracting parameters from the \acp{mpc} of \ac{rf} signals using a parametric channel estimation algorithm \cite{ShutWanJos:CSTA2013, HanFleRao:TSP2018,GreLeiWitFle:TWC2024}. 
%
A complicating factor in \ac{Mpslam} is measurement-origin uncertainty, i.e., the unknown association of measurements with features and their time-varying and unknown number\cite{LeiMeyHlaWitTufWin:J19, LeiGreWit:ICC2019, MenMeyBauWin:J19, LiLeiVenTuf:TWC2022}. State-of-the-art methods consider these challenges in their joint statistical model. In particular, \cite{LeiMeyHlaWitTufWin:J19, LeiGreWit:ICC2019, MenMeyBauWin:J19} use \ac{pda} and perform the \ac{spa} on a factor graph representing the underlying statistical model. Most of these methods rely on sampling techniques since the models for the aforementioned measurements are nonlinear \cite{WitMeiLeiSheGusTufHanDarMolConWin:J16, LeitingerJSAC2015, GentnerTWC2016, LeiMeyHlaWitTufWin:J19, LeiGreWit:ICC2019, MenMeyBauWin:J19}. Recently, \ac{Mpslam} was applied to data collected in indoor scenarios by radios with ultra-wide bandwidth or multiple antennas \cite{GentnerTWC2016,  LeiGreWit:ICC2019}. In \cite{WieVenWilLei:JAIF2023,WieVenWilWitLei:FUSION2024}, a method was developed to deal with non-ideal surfaces leading to possibly multiple reflections per VA. In \cite{LeiVenTeaMey:TSP2023} a novel \ac{Mpslam}
method is proposed that performs data fusion across multiple propagation path considering single-bounce and double-bounce reflections.
Similar approaches were proposed in \cite{Sharma2019,Brambilla2022} for simulaneous localization and tracking.

\subsection{Contributions}

In this paper, we introduce a Bayesian particle-based \ac{spa} for cooperative \ac{Mpslam} based on a factor graph that enables data fusion over different observations of map features (\acp{va}) by different \acp{mt} as well as cooperative MT-to-MT measurements using \ac{rf} signals. Our algorithm jointly performs \ac{pda} and sequential estimation of the states of the \acp{mt} and of \acp{pva} characterizing the map. The resulting \ac{spa} can infer \acp{va} by combining information across multiple \acp{mt} for each \ac{bs}. Furthermore, by considering \ac{mt}-to-\ac{mt} \ac{rf} signals, the \acp{spa} performs cooperative localization using \ac{pda}. In \cite{KimGraGaoBatKimWym:TWC2020} a similar approach was proposed based on the probability hypothesis density filter with a different map fusion scheme.
The key contributions are as follows.
\begin{itemize}
\item We model the exchange of information between different MTs, allowing for cooperative localization and data fusion of VAs over different MTs \cite{LeiVenTeaMey:TSP2023,WieLeiMeyWitTSIPN2022}.
\item We fully integrate an \ac{imu} as an additional sensor for each MT. This unlocks additional information for orientation and state transition estimation allowing to cope with complex trajectories.
\item We analyze the impact of \ac{va} data fusion and cooperative measurements in \ac{Mpslam} for \ac{mimo} and \ac{simo} systems using numerical simulation. In addition, we present the impact of using an IMU for the aforementioned cases.
\end{itemize}
Note that random variables are displayed in sans serif, upright fonts; their realizations in serif, italic fonts.

%% file: inputFiles/sigmod.tex
\section{Geometrical Relations}
At time step $n$, we consider $I$ \acp{mt} with unknown state $\V{x}_{i,n} = [\V{p}_{i,n} \iist \V{v}_{i,n} \iist \V{o}_{i,n}]^\text{T}$ at positions $\V{p}_{i,n} = [p_{x,i,n} \iist p_{y,i,n}]^\text{T}$ moving with velocity $\V{v}_{i,n} = [v_{x,i,n} \iist v_{y,i,n}]^\text{T}$ in the direction of $o_{i,n}$ where $i \in \{1, \dots,I\}$ and $J$ \acp{bs} with known positions $\V{p}_{1,\text{va}}^{(j)} \triangleq \V{p}_{\mathrm{bs}}^{(j)}, j \in \{1, \dots, J\}$. Each \ac{bs} $j$ has $N_{n}^{(j)} - 1$ specular reflections of radio signals at flat surfaces associated to it, modeled by \acp{va} with unknown positions $\V{p}_{l,\mathrm{va}}^{(j)} = \big[ {p}_{x,l,\mathrm{va}}^{(j)} \ist\ist\ist {p}_{y,l,\mathrm{va}}^{(j)} \big]^{\mathrm{T}} , l \in \{2, \dots, N_{n}^{(j)}\}$ (see \cite{LeiVenTeaMey:TSP2023}). By applying the image-source model \cite{Bor:JASA1984,LeiVenTeaMey:TSP2023}, a \ac{va} associated with a single-bounce path is the mirror image of $\V{p}_{\mathrm{bs}}^{(j)}$ at reflective surface $l$ given by
\vspace*{-1.5mm}  
\begin{align}
	\V{p}^{(j)}_{l,\mathrm{va}} &= \V{p}^{(j)}_{\mathrm{bs}} + 2\big( \V{u}_l^{\T}\V{e}_l - \V{u}_l^{\T}\V{p}^{(j)}_{\mathrm{bs}}\big)\V{u}_l \label{eq:VA1BPAequation}
	\\[-6mm]
	\nn
\end{align}
where $\V{u}_l$ is the normal vector of reflective surface $l$, and $\V{e}_l$ is an arbitrary point on the considered surface. 
The according point of reflection $\V{q}_{l,i,n}^{(j)}$ at the surface is given as
\vspace*{-2mm}
\begin{align}
\V{q}_{l,i,n}^{(j)} = \V{p}_{l, \text{va}}^{(j)} + \frac{(\V{p}_{\text{bs}}^{(j)} - \V{p}_{l, \text{va}}^{(j)})^\T \V{u}_l}{2(\V{p}_{i,n} - \V{p}_{l, \text{va}}^{(j)})^\T \V{u}_l} (\V{p}_{i,n} - \V{p}_{l, \text{va}}^{(j)})
\label{eq:reflectionPoint}\\[-7mm]\nn
\end{align}
and is needed to relate \ac{aod} of a specular reflection to the corresponding \ac{va}.
%
Both the \acp{mt} and all \acp{bs} are equipped with antenna arrays. The geometry of an antenna array is represented by its array element positions, which are defined for the \acp{bs} arrays by the distances $d^{(j,h)}$ and the angles $\psi^{(j,h)}$ relative to the \ac{bs} position $\V{p}_{\mathrm{bs}}^{(j)}$ (with known orientation $o_\text{bs}^{(j)}$), and for \ac{mt} array by distance $d^{(h)}$ and angle $\psi^{(h)}$ relative to the \ac{mt} position $\V{p}_{i,n}$ and unknown orientation $o_{i,n}$.
The radio signal (see Appendix~\ref{app:signal}) arrives at the receiver via the \ac{los} path as well as via \acp{mpc} originating from the reflection of surrounding objects.
The geometric relations of the \acp{mpc} parameters delay, \ac{aoa} and \ac{aod} w.r.t. MT $i$ and PVA $j$ are respectively given by ${\tau_\text{bs}}_{l,n}^{(j,i)} \triangleq d(\V{p}_{i,n}, \V{p}_{l,\text{va}}^{(j)})/c = \|\V{p}_{i,n} \!-\rmv \V{p}_{l,\text{va}}^{(j)}\|/c$,  $\theta^{(j,i)}_{l,n} \triangleq \angle(\V{p}_{i,n},\V{p}_{l,\text{va}}^{(j)})$ with $ \angle(\V{p}_{i,n},\V{p}_{l,\text{va}}^{(j)}) = \atantwo\big(p_{y,l,\text{va}}^{(j)}-p_{y,i,n},p_{x,l,\text{va}}^{(j)}-p_{x,i,n})$ and $\vartheta^{(j,i)}_{l,n} \triangleq \angle(\V{p}_{1,\text{va}}^{(j)},\V{p}_{i,n})$ for $l=1$ and $\vartheta^{(j,i)}_{l,n} \triangleq \angle(\V{q}_{l,i,n}^{(j)},\V{p}_{i,n})$ for $l>1$ (see Fig.~\ref{fig:scenario}).

\section{Parametric Channel Estimation: Measurements}  \label{sec:channel_estimation}

By applying at each time $ n $, a \ac{ceda} \cite{ShutWanJos:CSTA2013,HanFleRao:TSP2018,GreLeiWitFle:TWC2024} to the full {observed} discrete signal vector $\V{r}_{\text{bs}_n}^{(j,i)}$ or $\V{r}_{\text{mt}_n}^{(j,i)}$ (for details see Appendix~\ref{app:signal}) of all antenna elements at MT $i$ and \ac{pa} $j$ or MT $j$, one obtains a number of $M_n^{(j,i)}$ measurements denoted by ${\V{z}^{(j,i)}_{m,n}}$ with $m \in  \Set{M}_n^{(j)} \triangleq \{1,\,\dots\,,M_n^{(j)}\} $.
Each $\V{z}^{(j,i)}_{m,n} = [\zd \ \zaoa \ \zaod \ \zu]^\text{T}$ representing a potential \ac{mpc} parameter estimate, contains a delay measurement $\zd \rmv\rmv\in\rmv\rmv [0, \tau_\text{max}]$, an \ac{aoa} measurement $\zaoa \rmv\rmv\in\rmv\rmv [-\pi, \pi]$, an \ac{aod} measurement $\zaod \rmv\rmv\in\rmv\rmv [-\pi, \pi]$ and a normalized amplitude measurement $\zu \rmv\rmv\in\rmv\rmv [\gamma, \infty )$, where $\gamma$ is the detection threshold. Hence the \ac{ceda} compresses the information contained in $\V{r}_{n}^{(j,i)}$ into $\vspace*{-1mm}\V{z}^{(j,i)}_{n} = [{\bm{z}^{(j,i)\text{T}}_{1,n}}  \rmv \cdots  {\V{z}^{(j,i)\text{T}}_{M_n^{(j,i)},n}}]^\text{T}$
\vspace{1mm} which is used by the proposed algorithm as a noisy measurement.


%% file: inputFiles/sysmodslam.tex
\section{System Model}
\label{sec:systemModel}

At each time $n$, the states of the \acp{mt} are given by $\RV{x}_{i,n} = [\RV{p}_{i,n}^\T \ist \RV{v}_{i,n}^\T \ist \RV{o}_{i,n}]$, where $\RV{v}_{i,n}$ are the \ac{mt} velocities and $\RV{o}_{i,n}$ are the orientations. As in \cite{MeyKroWilLauHlaBraWin:J18, LeiMeyHlaWitTufWin:J19, LeiVenTeaMey:TSP2023}, we account for the unknown number of \acp{va} by introducing \acfp{pva} $k \in \mathcal{K}^{(j)}_n \triangleq \{1,\dots,\rv{K}_n^{(j)}\}$. The number $\rv{K}_n^{(j)}$ of \acp{pva} for each \ac{bs} $j$ is the
maximum possible number of actual \acp{va}, i.e., all \acp{va} that produced a measurement so far \cite{MeyKroWilLauHlaBraWin:J18,LeiMeyHlaWitTufWin:J19,LeiVenTeaMey:TSP2023} (where $\rv{K}_n^{(j)}$ increases
with time). The states of the \acp{pva} are given by $\RV{y}_{k,n}^{(j)} \rmv=\rmv \big[\RV{p}^{(j)\ist\T}_{k,\text{va}} \; \rv{u}^{(j)}_{k,n} \; \rv{r}^{(j)}_{k,n} \big ]^\T \rmv=\rmv \big[{\RV{x}}^{(j)\ist\T}_{k,n} \; \rv{r}^{(j)}_{k,n} \big ]^\T$, where $\rv{u}^{(j)}_{k,n}$ represents the normalized amplitude state of \ac{pva} $k$. The existence/nonexistence of \acp{pva} $k$ is modeled by the existence variable $\rv{r}_{k,n}^{(j)} \in \{0,1\}$ in the sense that \acp{pva} exists if $r_{k,n}^{(j)} = 1$. Formally, its states is considered even if
\acp{pva} $k$ is nonexistent, i.e., if $r_{k,n}^{(j)} = 0$. 
The states ${\RV{x}}^{(j)}_{k,n}$ of nonexistent \acp{pva} are irrelevant, hence \acp{pdf} defined for \ac{pva} states are of the form $f({\V{x}}^{(j)}_{k,n}, 0 )\rmv\rmv=\rmv\rmv f^{(j)}_{k,n} f_{\text{d}}({\V{x}}^{(j)}_{k,n})$, where $f_{\text{d}}({\V{x}}^{(j)}_{k,n})$ is an arbitrary ``dummy \ac{pdf}'' and $f^{(j)}_{k,n} \!\rmv\in [0,1]$ is a constant and can be interpreted as the probability of non-existence \cite{MeyKroWilLauHlaBraWin:J18, LeiMeyHlaWitTufWin:J19}. 
\subsection{BS-MT Measurement Model and New PVAs}\label{sec:measmodslam}

An existing \ac{pva} generates a measurement ${\RV{z}_\text{bs}}^{(j,i)}_{m,n}$ with $m \in  {\mathcal{M}_\text{bs}}^{(j,i)}_{n} \triangleq \{1,\dots,{\rv{M}_\text{bs}}^{(j,i)}_{n}\}$ with detection probability $ {p_{\mathrm{d}}}(\rv{u}^{(j)}_{k,n}) $. The single measurement \ac{lhf} is given by $ f({\V{z}_\text{bs}}^{(j,i)}_{m,n}|\V{x}_{i,n}, \V{y}^{(j)}_{k,n}) $, and it is assumed to be conditionally independent across the individual measurements within the vector ${\RV{z}_\text{bs}}^{(j,i)}_{m,n}$, thus it factorizes as
\begin{align}
f({\V{z}_\text{bs}}^{(j,i)}_{m,n}| \V{x}^{(j)}_{k,n},\V{x}_{i,n})	& = f(\zd | \V{p}_{i,n}, \V{x}^{(j)}_{k,n})  f(\zu | u^{(j)}_{k,n}) \nn \\
	& \hspace{-12mm} \times f(\zaoa | \V{x}_{i,n}, \V{x}^{(j)}_{k,n}) f(\zaod| \V{p}_{i,n}, \V{x}^{(j)}_{k,n}) 
	\label{eq:LHF_factorized}
\end{align}
where the individual \acp{lhf} of the distance, \ac{aoa} and \ac{aod} measurements are modeled by Gaussian \acp{pdf} $f_{\mathrm{N}}(x;\mu,\sigma^2)$ \cite{Kay1993}. More specifically, the \ac{lhf} of the distance measurement is given by
\begin{align}
\hspace{-3mm}	f(\zd| \V{p}_{i,n}, \V{x}^{(j)}_{k,n}) = \rmv f_{\mathrm{N}}(\zd; d(\V{p}_{i,n},\V{p}^{(j)}_{k,\text{va}}), \sigma_{\mathrm{d}}^2(u^{(j)}_{k,n}))
	\label{eq:LHF_dist}
\end{align}
where $ \sigma_{\mathrm{d}}^2({u}^{(j)}_{k,n}) = c^2/(8\pi^2 \beta_{\mathrm{bw}}^2 {u}_{k,n}^{(j)\ist2})$ is determined based on the Fisher information with $ \beta_{\mathrm{bw}}^2$ denoting the mean square bandwidth of $S^{(j)}_\text{bs}(f)$ \cite{WitrisalWCL2016, WilGreLeiMueWit:ACSSC2018,MenWymBauAbu:TWC2019,FacDeuKesWilColWitLeiSecWym:ICC2023}.
The \ac{lhf} of the \ac{aoa} and \ac{aod} measurements are given by\footnote{Note that a von Mises \ac{pdf} would model the angular distribution more accurately \cite{HanFleRao:TSP2018}. However, for reasonable \ac{aoa} variances, a Gaussian \ac{pdf} represents a sufficient approximation.}
\begin{align}
&\hspace{-10mm} f(\zaoa| \V{x}_{i,n}, \V{x}^{(j)}_{k,n}) \nn \\
&= f_{\mathrm{N}}(\zaoa; \angle( \V{p}_{i,n}, \V{p}^{(j)}_{k,\text{va}})- o_{i,n}, \sigma_{\theta}^2(u^{(j)}_{k,n}) ) 
	\label{eq:LHF_aAoA} \\
&\hspace{-10mm} f(\zaod | \V{p}_{i,n}, \V{x}^{(j)}_{k,n}) \nn \\
&= f_{\mathrm{N}}(\zaod; \angle( \V{p}_{i,n}, \V{p}^{(j)}_{k,\text{va}}), \sigma_{\vartheta}^2(u^{(j)}_{k,n}) ) 
	\label{eq:LHF_aAoD}
\end{align}
where $\sigma_{\mathrm{\theta}}^2({u}^{(j)}_{k,n})  = 1/\big(8\pi^2 {u}^{(j)\ist2}_{k,n} D^2\big(\angle(\V{p}_{i,n},\V{p}^{(j)}_{k,\text{va}})- {o}_i\big)$ and $\sigma_{\mathrm{\vartheta}}^2({u}^{(j)}_{k,n})  = 1/\big(8\pi^2 {u}^{(j)\ist2}_{k,n} D^2\big(\angle(\V{p}^{(j)}_{k,\text{va}},\V{p}_{i,n})\big)$ are determined based on the Fisher information with $D^2\big(\cdot\big)$ as normalized squared aperture of the antenna array \cite[Equation (24)]{WilGreLeiMueWit:ACSSC2018}.
The \ac{lhf} of the normalized amplitude measurements $ {u}^{(j)}_{k,n} \rmv>\rmv \gamma $ is modeled by a truncated Rician \ac{pdf} \cite[Ch.\,1.6.7]{BarShalomBook:Book2001}, \cite{LiLeiVenTuf:TWC2022}, i.e.,
\vspace*{-4mm}
\begin{align}
\hspace{-3mm} f(\zu | u^{(j)}_{k,n}) \rmv  \rmv = \rmv \rmv  f_{\mathrm{TR}}(\zu; u^{(j)}_{k,n}, \sigma_{\mathrm{u}}(u^{(j)}_{k,n}), {p_{\mathrm{d}}}(u^{(j)}_{k,n}); \gamma)
	\label{eq:LHF_normAmp}\\[-5mm]\nn
\end{align}
where the scale parameter $ \sigma_{\mathrm{u}}({u}^{(j)}_{k,n}) $ is determined based on the Fisher information given as $ \sigma_{\mathrm{u}}^2({u}^{(j)}_{k,n}) = \frac{1}{2} + \frac{1}{4 MH} {u}^{(j)\ist2}_{k,n} $. The variance $\sigma_{\mathrm{u}}^2({u}^{(j)}_{k,n})$ also considers for the measurement noise variance estimation of the \ac{ceda} (preprocessing) \cite{LiLeiVenTuf:TWC2022}. The $ {p_{\mathrm{d}}}({u}^{(j)}_{k,n}) $\cite{Kay1998,BarShalomBook:Book2001,LiLeiVenTuf:TWC2022} is directly related to the \ac{pva}'s visibility and the detection threshold $u_\text{de}$ of the \ac{ceda} (preprocessing) is directly connected to the likelihood model of the proposed \ac{Mpslam} (see \cite{LiLeiVenTuf:TWC2022}). \ac{Fa} measurements originating from the snapshot-based parametric channel estimator are assumed statistically independent of \ac{pva} states. These measurements are modeled by a Poisson process with mean number $ \mu_{\mathrm{fa}} $ and \ac{pdf} $ f_{\mathrm{fa}}({\V{z}_\text{bs}}^{(j,i)}_{m,n}) $, which is factorized as $ f_{\mathrm{fa}}({\V{z}_\text{bs}}^{(j,i)}_{m,n}) =  f_{\mathrm{fa}}(\zd) f_{\mathrm{fa}}(\zaoa)f_{\mathrm{fa}}(\zaod)f_{\mathrm{fa}}(\zu)$. The \ac{Fa} \acp{lhf} of the measurements corresponding to distance, \ac{aoa} and \ac{aod} are uniformly distributed on $ [0,d_{\mathrm{max}}] $, $ [-\pi,\pi)$ and $ [-\pi,\pi)$, respectively. The FA \ac{lhf} $ f_{\mathrm{fa}}(\zu) $ of the normalized amplitude is given by a truncated Rayleigh \ac{pdf} (see \cite{LiLeiVenTuf:TWC2022,VenLeiTerWit:TWC2023} for details).  

\textit{Newly detected \acp{va}} i.e., \acp{va} that generated a measurement for the first time, are modeled by a Poisson process with mean $\mu_{\mathrm{n}}$ and \ac{pdf} $f_{\mathrm{n}}(\overline{\V{x}}^{(j,i)}_{m,n}|\V{x}_{i,n})$.
Newly detected \acp{va} are represented by \textit{new \ac{pva} states} $\overline{\RV{y}}^{(j,i)}_{n,m}$, $m \rmv\rmv \in \{1,\dots,  {\rv{M}_\text{bs}}^{(j,i)}_{n} \}$ in our statistical model \cite{MeyKroWilLauHlaBraWin:J18, LeiMeyHlaWitTufWin:J19}. Each new \ac{pva} state corresponds to a measurement ${\RV{z}_\text{bs}}^{(j,i)}_{m,n}$; $\overline{r}_{m,n} \!=\! 1$ implies that measurement ${\RV{z}_\text{bs}}^{(j,i)}_{m,n}$ was generated by a newly detected \ac{va}. We denote by $\overline{\RV{y}}^{(j,i)}_n \triangleq \big [ \ist \overline{\RV{y}}^{(j,i)\ist\T}_{1,n} \ist\cdots\ist \overline{\RV{y}}^{(j,i)\ist \T}_{ {\rv{M}_\text{bs}}^{(j,i)}_{n} \rmv\rmv\rmv,n} \ist \big]^{\T}\rmv\rmv\rmv$, the joint vector of all new \ac{pva} states. Introducing new \ac{pva} for each measurement leads to a number of \ac{pva} states that grows with time $n$. Thus, to keep the proposed \ac{Mpslam} algorithm feasible, a sub-optimum pruning step is performed, removing \acp{pva} with a low probability of existence.

\subsection{Legacy PVAs and Sequential Update}

At time $n$, measurements are incorporated sequentially across \acp{mt} $i \rmv\in\rmv \{1,\dots,I\}$  \cite{MeyKroWilLauHlaBraWin:J18,LeiVenTeaMey:TSP2023}. Previously detected \acp{va}, i.e., \acp{va} that have been detected either at a previous time $n' \!<\rmv n$ or at the current time $n$ but at a previous \ac{mt} $i' \!<\rmv  i$, are represented by legacy \ac{pva} states $\underline{\RV{y}}_{k,n}^{(j)}$. New \acp{pva} become legacy \acp{pva} when the next measurements, either of the next \ac{mt} or at the next time instance, are taken into account. In particular, the \ac{va} represented by the new \ac{va} state $\overline{\RV{y}}^{(j,i')}_{m',n'}$ introduced due to measurement $m'$ of \ac{mt} $i'$ at time $n' \leq n$ is represented by the legacy \ac{pva} state $\underline{\RV{y}}^{(j)}_{k,n}$ at time $n$, with $k = \rv{K}^{(j)}_{n'-1} + \sum^{i' - 1}_{i'' = 1}  {\rv{M}_\text{bs}}^{(j,i'')}_{n} + m'$. The number of legacy \ac{pva} at time $n$, when the measurements of the next \ac{mt} $i$ are incorporated, is updated according to $\rv{K}^{(j,i)}_{n} = \rv{K}^{(j,i-1)}_{n-1} + {\rv{M}_\text{bs}}^{(j,i-1)}_{n}$, where $\rv{K}^{(j,1)}_{n} \rmv=\rmv \rv{K}^{(j)}_{n-1}$. Here, $\rv{K}^{(j,i)}_{n}$ is equal to the number of all measurements collected up to time $n$ and \ac{mt} $i \rmv-\rmv 1$. The vector of all legacy \ac{pva} states at time $n$ and up to \ac{mt} $i$ can now be written as $\underline{\RV{y}}^{(j,i)}_{n} = \big[\underline{\RV{y}}^{(j,i-1) \T}_{n} \ist\ist\ist \overline{\RV{y}}^{(j,i-1) \T}_{n} \big]^{\T}\rmv\rmv$ and the current vector containing all states as ${\RV{y}}^{(j,i-1)}_{n} = \big[\underline{\RV{y}}^{(j,i-1) \T}_{n} \ist\ist\ist \overline{\RV{y}}^{(j,i-1) \T}_{n} \big]^{\T}\rmv\rmv$ where $\underline{\RV{y}}^{(j,1)}_n \rmv\triangleq\rmv \big[ \underline{\RV{y}}^{(j)\ist\T}_{1,n} \rmv\cdots\ist \underline{\RV{y}}^{(j)\ist \T}_{\rv{K}^{(j)}_{n\rmv-\rmv1},n} \big]^{\T}\rmv\rmv$ is the vector of all legacy \ac{pva} states before any measurements at time $n$ have been incorporated. After the measurements of all \acp{mt} $i \in \{1,\dots,I\}$ have been incorporated at time $n$, the total number of \ac{pva} states is $\rv{K}^{(j)}_n = \rv{K}^{(j)}_{n-1} + \sum_{i=1}^{I}  {\rv{M}_\text{bs}}^{(j,i)}_{n} = \rv{K}^{(j,I)}_{n} +{\rv{M}_\text{bs}}^{(j,I)}_{n}$
and the vector of all \ac{pva} states at time $n$ is given by $\RV{y}^{(j)}_{n} \!=\rmv \big[\underline{\RV{y}}^{(j,I)\ist \T}_{n} \ist\ist \overline{\RV{y}}^{(j,I) \T}_{n} \big]^\T$.

\subsection{State Evolution}
Legacy \acp{pva} states $\underline{\RV{y}}^{(j)}_{k,n}$ and the \ac{mt} states $\RV{x}_{i,n}$ are assumed to evolve independently across time according to state-transition \acp{pdf} $f\big(\underline{\V{y}}^{(j)}_{k,n} \big| \V{y}_{k, n-1}\big)$ and $f(\V{x}_{i,n}|\V{x}_{i,n-1},\V{c}_{i,n-1})$, respectively, where $\V{c}_{i,n-1}$ is a control term. If \ac{pva} $k$ exists at time $n \rmv-\! 1$, i.e., $r^{(j)}_{k,n-1} \!=\! 1$, it either disappears, i.e., $\overline{r}^{(j)}_{k,n} \!=\rmv 0$, or survives, i.e., $\overline{r}^{(j)}_{k,n} \!=\! 1$; in the latter case, it becomes a legacy \ac{pva} at time $n$. The probability of survival is denoted by $p_\mathrm{s}$. In case of survival, its position remains unchanged, i.e., the state-transition \ac{pdf} of the \ac{va} positions $\underline{\RV{p}}^{(j)}_{k,\mathrm{va}}$ is given by $f\big(\underline{\V{p}}^{(j)}_{k,\mathrm{va}}  \ist \big| \ist \V{p}^{(j)}_{k,\mathrm{va}} \big) = \delta \big(\underline{\V{p}}^{(j)}_{k,\mathrm{va}} \rmv - \ist \V{p}^{(j)}_{k,\mathrm{va}} \big)$ and for the normalized amplitude $\rv{u}^{(j)}_{k,n}$ is given by $f\big(\underline{u}^{(j)}_{k,n} \ist \big| \ist u^{(j)}_{k,n-1} \big) $.
Therefore, $f\big(\underline{\V{x}}^{(j)}_{k}\rmv,\underline{r}^{(j)}_{k,n} \ist \big| \ist \V{x}^{(j)}_{k} , r^{(j)}_{k,n-1} \big) $ for $r^{(j)}_{k,n-1} \rmv=\rmv 1$ is obtained as  
\vspace*{-1mm}
\begin{align}
	f\big(\underline{\V{x}}^{(j)}_{k},\underline{r}^{(j)}_{k,n} \ist \big| \ist \V{x}^{(j)}_{k} , r^{(j)}_{k,n-1} = 1\big) & \nn \\
	& \hspace{-40mm} =\rmv\begin{cases} 
		(1 \!-\rmv p_\mathrm{s}) \ist f_\text{d}\big(\underline{\V{p}}^{(j)}_{k,\mathrm{va}} \big) , &\!\!\! \underline{r}^{(j)}_{k,n} \!=\rmv 0 \\[0mm]
		p_\mathrm{s} \ist\ist f\big(\underline{u}^{(j)}_{k,n} \ist \big| \ist u^{(j)}_{k,n-1} \big)  \delta \big(\underline{\V{p}}^{(j)}_{k,\mathrm{va}} \rmv - \ist \V{p}^{(j)}_{k,\mathrm{va}} \big) , &\!\!\! \underline{r}^{(j)}_{k,n} \!=\! 1.
	\end{cases}\label{eq:stmpvarone}
\end{align}
If \ac{va} $k$ does not exist at time $n \rmv-\! 1$, i.e., $r^{(j)}_{k,n-1} \!=\! 0$, it cannot exist as a legacy \ac{pva} at time $n$ either, thus we get
\vspace*{-1mm}
\begin{align}
	&f\big(\underline{\V{x}}^{(j)}_{k},\underline{r}^{(j)}_{k,n} \big| \V{x}^{(j)}_{k} , r^{(j)}_{k,n-1} \rmv\rmv= \rmv\rmv 0\big) =\rmv\begin{cases} 
		f_\text{d}\big( \underline{\V{x}}^{(j)}_{k} \big) , &\!\!\! \underline{r}^{(j)}_{k,n} \!=\rmv 0 \\[0mm]
		0 , &\!\!\! \underline{r}^{(j)}_{k,n} \!=\! 1.
	\end{cases}\label{eq:stmpvarzero}
\end{align}
For $i \geq 2$, we also define $f^{(i)}\big(\underline{\V{y}}^{(j,i)}_{k,n} \big| {\V{y}}^{(j,i-1)}_{k, n}\big)$ as
\vspace*{-1mm}
\begin{align}
f^{(i)}\big(\underline{\V{x}}^{(j,i)}_{k},\underline{r}^{(j,i)}_{k,n} \ist \big| {\ist {\V{x}}^{(j,i-1)}_{k} , {r}^{(j,i-1)}_{k,n} = 1}\big) & \nn \\
& \hspace{-30mm} =\rmv\begin{cases} 
		f_\text{d}\big(\underline{\V{x}}^{(j,i)}_{k} \big) , &\!\!\! \underline{r}^{(j,i)}_{k,n} \!=\rmv 0 \\[0mm]
		\delta \big(\underline{\V{x}}^{(j,i)}_{k} \rmv - \ist {\V{x}}^{(j,i-1)}_{k} \big) , &\!\!\! \underline{r}^{(j,i)}_{k,n} \!=\! 1
	\end{cases}\label{eq:stmpvaonepas}
\end{align}
and
\vspace*{-1mm}
\begin{align}
f^{(i)}\big(\underline{\V{x}}^{(j,i)}_{k},\underline{r}^{(j,i)}_{k,n} \big| {{\V{x}}^{(j,i-1)}_{k} , {r}^{(j,i-1)}_{k,n} \rmv\rmv= \rmv\rmv 0}\big) & \nn \\ & \hspace{-20mm} =\rmv\begin{cases} 
		f_\text{d}\big( \underline{\V{x}}^{(j,i)}_{k} \big) , &\!\!\! \underline{r}^{(j,i)}_{k,n} \!=\rmv 0 \\[0mm]
		0 , &\!\!\! \underline{r}^{(j,i)}_{k,n} \!=\! 1.
	\end{cases}\label{eq:stmpvazeropas}
\end{align}
It it assumed that at time $n \rmv=\rmv 0$ the initial prior \ac{pdf} $f\big(\V{y}^{(j)}_{k,0} \big)$, $k = \big\{1,\dots,K^{(j)}_0\big\}$ and  $f(\V{x}_{i,0})$ are known. All (legacy and new) \ac{pva} states and all \ac{mt} states up to time $n$ are denoted as $\RV{y}_{n} \!=\rmv \big[\RV{y}^{(1)\ist \T}_{n} \cdots\ist \RV{y}^{(J) \T}_{n} \big]^\T$ and $\RV{y}_{0:n} \triangleq \big[\RV{y}^{\T}_{0} \cdots\ist \RV{y}^{\T}_{n} \big]^{\T}\!$ and $\RV{x}_{n} = [\RV{x}_{1,n}^\T \ist\cdots \RV{x}_{I_{\text{MT}},n}^\T]$ and $\RV{x}_{0:n} = [\RV{x}_{0}^\T \ist\cdots \RV{x}_{n}^\T]$, respectively.

\subsection{MT-MT Measurement Model}\label{eq:lhfmodelMTMT}

An existing \ac{los} between \ac{mt} $i$ and \ac{mt} $i'$ generates a measurement ${\RV{z}_\text{co}}^{(i,i')}_{m,n} \in {\mathcal{M}_\text{co}}^{(i,i')}_{n} \triangleq \{1,\dots,{\rv{M}_\text{co}}^{(j,i)}_{n}\} $. The single measurement \ac{lhf} is given by $ f({\V{z}_\text{co}}^{(i,i')}_{m,n}|\V{x}_{i,n}, \V{x}_{i',n}) $, and it is assumed to be conditionally independent across the individual measurements within the vector ${\RV{z}_\text{co}}^{(i,i')}_{m,n}$. Since the orientation of both MTs is a nuisance parameter, the angular measurements do not provide additional information in that case. Hence, we only consider distance measurements and neglect angular measurements for cooperation. The \acp{lhf} is given as
\begin{align}
	f({\V{z}_\text{co}}^{(i,i')}_{m,n}|\V{x}_{i,n},\V{x}_{i',n}) & = f({z_{\mathrm{co,d}}}^{(i,i')}_{m,n} | \V{p}_{i,n}, \V{p}_{i',n}) \label{eq:LHF_factorized}
\end{align}

\subsection{\ac{imu} Measurement Model}\label{sec:nav}
At each time step $n$, an \ac{imu}, consisting of an accelerometer, a gyroscope and a magnetometer, provides each \ac{mt} with the corresponding \ac{imu} measurements $\V{z}_{\text{acc}_{i,n}}, \V{z}_{\text{gyr}_{i,n}}$ and $\V{z}_{\text{mag}_{i,n}}$, respectively. The normalized acceleration measurement is defined as $\tilde{\V{z}}_{\text{acc}_{i,n}} = \V{z}_{\text{acc}_{i,n}} / \|\V{z}_{\text{acc}_{i,n}}\|$. Therefor, we define $\V{z}_{\text{IMU}_{i,n}} = [\tilde{\V{z}}_{\text{acc}_{i,n}}^\T \V{z}_{\text{gyr}_{i,n}}^\T \tilde{\V{z}}_{\text{mag}_{i,n}}^\T]^\T$. The \ac{imu} \ac{lhf} is given as
\vspace{-2mm}
\begin{align}
f(\V{z}_{\text{IMU}_{i,n}} | \V{x}_{i,n'}) & = f_{\mathrm{N}}(\V{z}_{\text{IMU}_{i,n}}, \V{o}_{i,n}, \Sigma_i) \label{eq:lhf_nav}
\end{align} 
where $\Sigma_i$ is a blockdiagonal covariance matrix. For the orientation state, a quaternion representation is used and the mapping of the \ac{imu} measurements to the quaternion state is done similarly to \cite{madgwick2010efficient}.

%
\begin{figure*}[!b]
\begin{align}
f\big( \V{y}_{0:n}, \V{x}_{0:n}, \underline{\V{a}}_{1:n},\overline{\V{a}}_{1:n},\V{b}_{1:n} \ist \big| \ist \V{z}_{1:n}\big) & \propto\underbrace{\bigg( \prod^{I}_{i = 1}\ist f(\V{x}_{i,0}) \ist \prod^{K_0}_{k = 1} \ist f(\V{y}_{k,0})  \bigg)}_{\footnotesize \text{MT and VA states initial prior PDFs}} \prod^{n}_{n' = 1}  \underbrace{\bigg( \prod^{I}_{i = 1}\ist f\big(\V{x}_{i,n'} | \V{x}_{i,n'-1},\V{c}_{i,n'-1}\big) f\big(\V{z}_{\text{IMU}_{i,n'}} | \V{x}_{i,n'}\big)}_{\footnotesize \text{MT state prediction and IMU measurement}} \nn \\
 & \hspace{-50mm} \times \underbrace{\prod^{I}_{i' = 1,i\neq i'}\ist \ist  g_\text{co}\big(\V{x}_{i,n'},\V{x}_{i',n'}, b^{(i,i')}_{n'};{\V{z}_\text{co}}_{n'}\big) \bigg)}_{\footnotesize \text{MT cooperative related factors}} \underbrace{\bigg(  \prod^{J}_{j = 1}\prod^{K^{(j)}_{n'\rmv-\rmv1}}_{k' = 1} f \big(\underline{\V{y}}^{(j)}_{k'\rmv\rmv,n'} | \V{y}^{(j)}_{k'\rmv\rmv,n'-1}\big) \rmv \bigg)}_{\footnotesize \text{Legacy PVA states prediction}} \underbrace{ \bigg(\prod^{J}_{j = 1} \ist\prod^{I}_{i = 2}\ist \prod^{K^{(j,i)}_{n'}}_{k = 1}\ist f^{(i)}\big(\underline{\V{y}}^{(j,i)}_{k,n'} \big| \underline{\V{y}}^{(j,i-1)}_{k, n'}\big)\bigg)}_{\footnotesize  \text{Legacy PVA states transition factors}}  \nn\\[-1mm]
 & \hspace{-50mm} \times \underbrace{\prod^{J}_{j = 1}\prod^{I}_{i = 1} \prod^{K^{(i)}_{n'}}_{k = 1} \ist \underline{q}_\text{BS}\big( \underline{\V{y}}^{(j,i)}_{k,n'}\rmv, \underline{a}^{(j,i)}_{k,n'},  \V{x}_{i,n'} ; {\V{z}_\text{bs}}^{(j)}_{i,n'} \big)\prod^{{M_\text{bs}}^{(j,i)}_{n'}}_{m = 1} \rmv\rmv \Psi\big(\underline{a}^{(j,i)}_{k,n'},\overline{a}^{(j,i)}_{m,n'} \big)}_{\footnotesize  \text{Legacy PVA states related factors}} \underbrace{ \prod^{{M_\text{bs}}^{(j,i)}_{n'}}_{m = 1} \overline{q}_\text{BS}\big( \overline{\V{y}}^{(j,i)}_{m,n'}, \overline{a}^{(j,i)}_{m,n'}, \V{x}_{n'} ; {\V{z}_\text{bs}}^{(j)}_{i,n'} \big) \rmv }_{\footnotesize  \text{New PVA states prior PDF and related factors}}\label{eq:jointpostpdf}
\end{align}
\end{figure*}

\subsection{Data Association}
The data association between MTs and MVAs as well as MTs and MTs is described in Appendix~\ref{app:dataassoc}.

\subsection{Joint Posterior PDF and Factor Graph}\label{sec:postpdfandfg}
Using Bayes’ rule and independence assumptions related to the state-transition \acp{pdf}, the prior \acp{pdf}, and the likelihood model (for details please see \cite{MeyKroWilLauHlaBraWin:J18, LeiMeyHlaWitTufWin:J19, LeiVenTeaMey:TSP2023}), and for fixed (observed) measurements $\V{z}_{1:n}$ (the numbers of measurements ${M_\text{bs}}^{(j,i)}_{n}$ and ${M_\text{co}}^{(i,i')}_{n}$ are fixed and not random anymore) the joint posterior \ac{pdf} of $\RV{y}_{0:n}$, $\RV{x}_{0:n}$, $\underline{\RV{a}}_{1:n}$, $\overline{\RV{a}}_{1:n}$ and $\RV{b}_{1:n}$ conditioned on $\V{z}_{1:n}$ is given by \eqref{eq:jointpostpdf} where we introduced the functions $g_\text{co}\big(\V{x}_{i,n},\V{x}_{i',n}, b^{(i,i')}_{n};{\V{z}_\text{co}}^{(i,i')}_{n}\big)$, $\underline{q}_\text{BS}\big( \underline{\V{y}}^{(j,i)}_{k,n},\underline{a}^{(j,i)}_{k,n},  \V{x}_{i,n} ; {\V{z}_\text{bs}}^{(j)}_{i,n} \big)$, and $\overline{q}_\text{BS}\big( \overline{\V{y}}^{(j,i)}_{m,n},$ $ \overline{a}^{(j,i)}_{m,n}, \V{x}_{i,n} ; {\V{z}_\text{bs}}^{(j)}_{i,n} \big)$ that will be discussed next. 

The \textit{pseudo likelihood functions} $g_\text{co}\big(\V{x}_{i,n},\V{x}_{i',n}, b^{(i,i')}_{n};{\V{z}_\text{co}}_{n}\big)$ and $\underline{q}_\text{BS}\big( \underline{\V{y}}^{(j,i)}_{k,n},\underline{a}^{(j,i)}_{k,n},  \V{x}_{i,n} ; {\V{z}_\text{bs}}^{(j)}_{i,n} \big)$ are respectively given for $(i,i') \in \mathcal{C}^{(i,'i)}_{n}$ by
\begin{align}
	g_\text{co}\big(\V{x}_{i,n},\V{x}_{i',n}, b^{(i,i')}_{n};{\V{z}_\text{co}}^{(i,i')}_{n}\big)  & \nn \\
	& \hspace{-45mm} \triangleq \begin{cases}
		\displaystyle \ist \frac{ 	p^{(j)}_{\mathrm{d}}(u_{i,n})  f\big( {\V{z}_\text{co}}^{(i,i')}_{m,n} \big|\ist \V{x}_{i,n},\V{x}_{i',n} \big)}{ \mu_{\mathrm{fa}} \ist f_{\mathrm{fa}}\big( {\V{z}_\text{co}}^{(i,i')}_{m,n} \big)} \ist, 
		& \!\!\rmv b^{(i,i')}_{n} \!=\rmv m \rmv\in\rmv {\Set{M}_\text{co}}^{(i,i')}_{n}\\[3.5mm]
		1 \!-\rmv 	p^{(j)}_{\mathrm{d}}(\V{p}_{i,n},\V{p}_{i',n}) \ist, & \!\!\rmv b^{(i,i')}_{n} \!=\rmv 0   
	\end{cases} \label{eq:cooplhf}
\end{align}
for \ac{bs} $j$ and \ac{mt} $i$ by
\begin{align}
\underline{q}_\text{BS}\big( \underline{\V{x}}^{(j,i)}_{k,n},\underline{r}^{(j,i)}_{k,n}=1,\underline{a}^{(j,i)}_{k,n},  \V{x}_{i,n} ; {\V{z}_\text{bs}}^{(j,i)}_{n} \big) & \nn \\
& \hspace{-60mm} \triangleq \begin{cases}
		\displaystyle \ist \frac{ 	p^{(j)}_{\mathrm{d}}(\underline{u}^{(j,i)}_{k,n})  f\big( {\V{z}_\text{bs}}^{(j,i)}_{m,n} \big|\ist \V{x}_{i,n}, \underline{\V{x}}^{(j,i)}_{k,n} \big)}{ \mu_{\mathrm{fa}} \ist f_{\mathrm{fa}}\big(  f\big( {\V{z}_\text{bs}}^{(j,i)}_{m,n} \big)} \ist, 
		& \!\!\rmv \underline{a}^{(j,i)}_{k,n} \!=\rmv m \rmv\in\rmv{\Set{M}_\text{bs}}^{(j,i)}_{n}\\[3.5mm]
		1 \!-\rmv 	p^{(j)}_{\mathrm{d}}(\underline{u}^{(j,i)}_{k,n}) \ist, & \!\!\rmv \underline{a}^{(j,i)}_{k,n} \!=\rmv 0   
	\end{cases} \label{eq:bslhf}
\end{align}
and $\underline{q}_\text{BS}\big( \underline{\V{x}}^{(j,i)}_{k,n},\underline{r}^{(j,i)}_{k,n}=0,\underline{a}^{(j,i)}_{k,n},  \V{x}_{i,n} ; {\V{z}_\text{bs}}^{(j,i)}_{n} \big)   \rmv\triangleq\rmv {\delta_{\underline{a}^{(j,i)}_{k,n}}}$.
The \textit{pseudo likelihood functions} related to new \acp{pva}  $\overline{q}_\text{BS}\big( \overline{\V{y}}^{(j,i)}_{m,n},$ $ \overline{a}^{(j,i)}_{m,n}, \V{x}_{i,n} ; {\V{z}_\text{bs}}^{(j)}_{i,n} \big)$ is given by
\begin{align}
\overline{q}_\text{BS}\big( \overline{\V{x}}^{(j,i)}_{m,n},\overline{r}^{(j,i)}_{m,n}=1, \overline{a}^{(j,i)}_{m,n}, \V{x}_{i,n} ; {\V{z}_\text{bs}}^{(j)}_{i,n} \big) \nn \\
& \hspace{-45mm} \triangleq \begin{cases}
		0    \ist, 
		& \hspace{-1mm} \overline{a}^{(j,i)}_{m,n}  \rmv\rmv\in\rmv \Set{K}_{n}^{(j,i)} \\[1mm] %
		\frac{ \mu_{\mathrm{n}}f_{\mathrm{n}}(\overline{\V{x}}^{(j,i)}_{m,n} \ist | \ist\V{x}_{i,n}) f({\V{z}_\text{bs}}^{(j,i)}_{m,n} \ist | \V{x}_{i,n}, \overline{\V{x}}^{(j,i)}_{m,n})}{\mu_{\mathrm{fa}}  f_{\mathrm{fa}}({\V{z}_\text{bs}}^{(j,i)}_{m,n} )} \ist,  & \hspace{-1mm} \overline{a}^{(j,i)}_{m,n} \rmv=\rmv 0  
	\end{cases}  \label{eq:factorvNewPVAs}
\end{align}
and $\overline{q}_\text{BS}\big( \overline{\V{x}}^{(j,i)}_{m,n},\overline{r}^{(j,i)}_{m,n}=0, \overline{a}^{(j,i)}_{m,n}, \V{x}_{i,n} ; {\V{z}_\text{bs}}^{(j)}_{i,n} \big) \rmv\triangleq\rmv 	f_\text{d}\big(\overline{\V{x}}^{(j,i)}_{m,n}\big)$, respectively. 

\subsection{Confirmation of PVAs and State Estimation}\label{sec:probFormulation}

We aim to estimate the \ac{mt} states $\RV{x}_{i,n}$ using all available measurements $\V{z}_{1:n}$ up to time $n$. In particular, we calculate an estimate $\hat{\V{x}}_{i,n}$ by using the \ac{mmse} estimator \cite[Ch.~4]{Kay1993}
\vspace*{-1mm}
\begin{equation}
	\hat{\V{x}}_{i,n} \ist\triangleq\, \int \rmv \V{x}_n \ist f( \V{x}_{i,n}|\V{z}_{1:n}) \ist \mathrm{d} \V{x}_{i,n} \rmv. \label{eq:MMSEMTs}
	\vspace*{-1mm}
\end{equation}
Estimating the positions $\RV{p}^{(j)}_{k,\mathrm{va}}$ of the detected \acp{pva} $k \!\in\! \{ 1,\dots,K^{(j)}_n \}$ rely on the marginal posterior existence probabilities $p(r^{(j)}_{k,n} \!=\! 1|\V{z}_{1:n}) = \int f(\V{p}^{(j)}_{k,\mathrm{va}} , r^{(j)}_{k,n} \!=\! 1| \V{z}_{1:n} ) \mathrm{d}\V{p}^{(j)}_{k,\mathrm{va}}$ and the marginal posterior \acp{pdf} $f(\V{p}^{(j)}_{k,\mathrm{va}} | r^{(j)}_{k,n} \!=\! 1, \V{z}_{1:n} ) \rmv\rmv=\rmv\rmv f(\V{p}^{(j)}_{k,\mathrm{va}} , r^{(j)}_{k,n} \!=\! 1| \V{z}_{1:n} )/p(r^{(j)}_{k,n}  \!=\! 1|\V{z}_{1:n})$. A \ac{pva} $k$ is declared to exist $p(r^{(j)}_{k,n} \!=\! 1|\V{z}_{1:n}) > p_{\mathrm{cf}}$, where $p_{\mathrm{cf}}$ is a confirmation threshold \cite{Kay1998}.  To avoid that the number of \acp{pva} states grows indefinitely, \acp{pva} states with $p(r^{(j)}_{k,n} \rmv \rmv=\rmv \rmv 1|\V{z}_{1:n}) < p_{\text{pr}}$ are removed from the state space (``pruned''). For existing \acp{pva}, estimates of it's position $\RV{p}^{(j)}_{k,\mathrm{va}}$ are calculated by the \acp{mmse}, i.e.,
\vspace*{-1mm}
\begin{align}
	\hat{\V{p}}^{(j)}_{k,\mathrm{va}} &\triangleq \int \rmv \V{p}^{(j)}_{k,\mathrm{va}}  \ist\ist f(\V{p}^{(j)}_{k,\mathrm{va}} \ist | \ist r^{(j)}_{k,n} \!=\! 1, \V{z}_{1:n}) \ist\ist \mathrm{d} \V{p}^{(j)}_{k,\mathrm{va}}  \label{eq:MMSEpva}\vspace*{-1mm}
\end{align}
The calculation of $f( \V{x}_{i,n}|\V{z}_{1:n})$, $p(r^{(j)}_{k,n}|\V{z}_{1:n})$, and $f(\V{p}^{(j)}_{k,\mathrm{va}} \ist | \ist r^{(j)}_{k,n}, \V{z}_{1:n})$ from the joint posterior $f( \V{y}_{0:n}, \V{x}_{0:n}, \underline{\V{a}}_{1:n},\overline{\V{a}}_{1:n},\V{b}_{1:n} \ist \big| \ist \V{z}_{1:n})$ by direct marginalization is not feasible. By performing sequential message passing using the SPA rules on a \ac{fg}\cite{KscFreLoe:TIT2001, MeyHliHla:TSPIN2016, MeyBraWilHla:J17, LeiMeyHlaWitTufWin:J19}, approximations (``beliefs'') $\tilde{f}\big(\V{x}_{i,n} \big)$ and $\tilde{f}^{(j)}_k\big(\V{y}^{(j)}_{k,n} \big)$ of the marginal posterior \acp{pdf} $f( \V{x}_{i,n}|\V{z}_{1:n})$, and $f(\V{p}^{(j)}_{k,\mathrm{va}} \ist | \ist r^{(j)}_{k,n}, \V{z}_{1:n})$ and the marginal \ac{pmf} $p(r^{(j)}_{k,n}|\V{z}_{1:n})$ can be obtained efficiently for the \ac{mt} states as well as all legacy and new \acp{pva} states $k \in \Set{K}^{(j)}_n$.

%% file: inputFiles/scheduling_and_MP.tex
\section{Scheduling and Message Passing}
The factorization in \eqref{eq:jointpostpdf} is represented by a \ac{fg}. The explicit scheduling of the message passing is directly encoded in the \ac{fg} and is mentioned in what follows.

\subsection{Prediction Step for \acp{pva}, \acp{mt} and \ac{imu} Update} \label{sec:pred_pva}
The prediction message for legacy \acp{pva} is based on \eqref{eq:stmpvarone} and \eqref{eq:stmpvarzero}. For each \ac{mt}, a state transition is performed according to $f(\V{x}_{i,n}|\V{x}_{i,n-1},\V{c}_{i,n-1})$. Based on the predicted \ac{mt} state and \eqref{eq:lhf_nav}, the orientation state of the stacked \ac{mt} state is updated and used in further calculations. The control term $\V{c}_{i,n-1}$ is given by transforming $\V{z}_{\text{acc}_{i,n}}$ from the MT frame to the navigation frame using the orientation state \cite{madgwick2010efficient,HolDijLuiSch:ICUW2009}.

\subsection{Sequential Update based on \acp{mt}}
The following calculations are performed for all legacy \acp{pva} and for all new \acp{pva} for all \acp{bs} $j \in \{1, . . . , J\}$ in parallel and for each \ac{mt} $i$ in a sequential manner. For the \ac{mt} states calculations are performed sequentially for each \acp{bs} $j$.
\subsubsection{Transition to current legacy PVAs}
For $i=1$, the messages representing the current beliefs of the legacy PVA states are determined based on Section~\ref{sec:pred_pva}. For $i>1$, the messages are determined by using \eqref{eq:stmpvaonepas} and \eqref{eq:stmpvazeropas}, meaning that new PVAs for MT $i$ become legacy PVAs for MT $i+1$.
\subsubsection{Measurement Evaluation for Legacy PVAs}
This is described by the messages passed from the factor node
$\underline{q}_\text{BS}(\cdot)$ of single PVAs to the feature-oriented association variables $\underline{a}^{(j,i)}_{k,n}$.
\subsubsection{Measurement Evaluation for New PVAs}
This is described by the messages passed from the factor node
$\overline{q}_\text{BS}(\cdot)$ of single PVAs to the feature-oriented association variables $\overline{a}^{(j,i)}_{k,n}$.
\subsubsection{Iterative Data Association}
These messages are obtained by performing loopy \ac{bp} based on the measurement evaluation messages. It has to be done to associate measurements to PVAs and cooperating MTs. It is implemented similar to \cite{LeiVenTeaMey:TSP2023}.
\subsubsection{Measurement Update}
The following measurement update steps are performed sequentially. At first, the MT states are updated by using only measurements associated to legacy PVAs. Afterwards the legacy PVA $(j,i)$ is updated by the MT state $(j-1,i)$. The update for new PVAs is done in a similar manner. If all MTs are considered, the belief of the PVAs is determined as $\tilde{f}(\V{y}_{k,n}^{(j)}) \triangleq \tilde{f}(\V{y}_{k,n}^{(j,I-1)})$. At last, the cooperative updates of MTs are performed. Note that they have no direct impact on the PVA updates at time $n$ but only indirectly via the state transition at $n+1$.

%% file: inputFiles/simulation_results.tex
\def\figureDataH{0.25\columnwidth}
\def\figureDataW{0.55\columnwidth}
\begin{figure}[!t]
	\centering
	\subfloat{\includegraphics{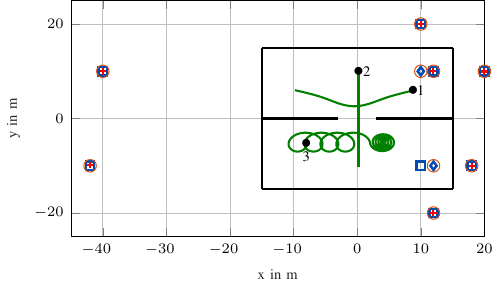}}\\
	\subfloat{\includegraphics{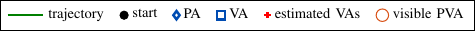}\hfill}\\
	\caption{Considered scenario for performance evaluation. The start of each trajectory is indicated by a black dot and the visibility is determined at the end of the trajectories. Note that not all VAs are shown.}
	\label{fig:floorplan}
\end{figure}
\begin{figure*}[!t]
	\centering
	\subfloat[]{\hspace{0mm}\includegraphics{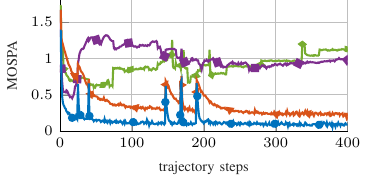}\label{fig:mospa}}
	\subfloat[]{\hspace{-3mm}\includegraphics{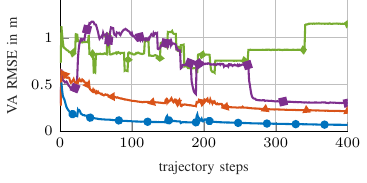}\label{fig:VA_RMSE}}
	\subfloat[]{\includegraphics{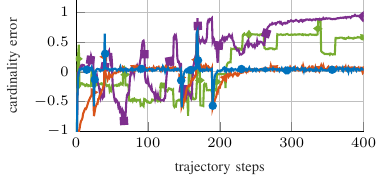}\label{fig:cardError}} \\
	\hspace*{0cm}
	\subfloat{\includegraphics{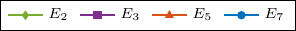}\hfill}\\
	\caption{Comparison of different experimental settings. (a) shows the MOSPA, (b) the RMSE of the position errors of estimated VAs that were correctly associated to a true VA and (c) shows the cardinaltiy error. The results are averaged over all simulation runs and estimated VAs. The legend description can be found in Table~\ref{tb:settings}.}
	\label{fig:results}
\end{figure*}
\begin{figure}[!t]
	\centering
	\subfloat{\includegraphics{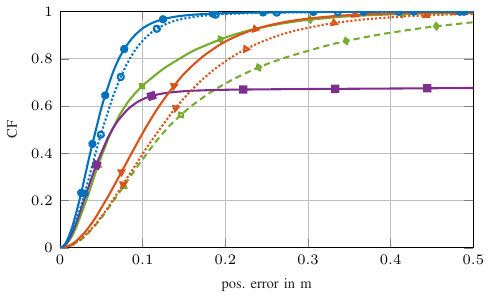}} \\
	\hspace*{-0mm}
	\subfloat{\includegraphics{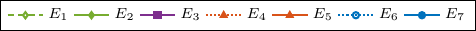}\hfill}\\
	\caption{Cumulative frequency of MT position errors for different settings.The legend description can be found in Table~\ref{tb:settings}.}
	\label{fig:cdf}
\end{figure}
\section{Numerical Results}
\label{sec:results}
We consider an indoor scenario shown in Fig.~\ref{fig:floorplan}. 
The scenario consists of two \acp{bs} and five reflective surfaces, i.e., $5$ \acp{va} per BS. Three \acp{mt} move along tracks which are observed for $400$ time instances $n$ with observation period $\Delta T = 1\,$s. 
Measurements are generated according to the proposed system model in Section~\ref{sec:systemModel}. The \ac{snr} is set to {$40\,\mathrm{dB}$} at an \ac{los} distance of $1\,$m. The amplitudes of the main-components (\ac{los} component and the \acp{mpc}) are calculated using a free-space path loss model and an additional attenuation of {$3\,\mathrm{dB}$} for each reflection at a surface.
We define $p_\text{d} = 0.98$. In addition false positive measurements are generated according to the model in Section~\ref{sec:systemModel} with a mean number of $\mu_\text{fp} = 5$. For the calculation of the measurement variances, we assume a $3\text{-}\mathrm{dB}$ system bandwidth of $B=500\,\mathrm{MHz}$ with some known transmit signal spectrum at a carrier frequency of $f_\mathrm{c} = 6\,\mathrm{GHz}$. The arrays employed at \acp{mt} and \acp{pa} are of identical geometry with $H = H^{(j)}=4$ antenna elements in (known) \ac{ura}-configuration spaced at $\lambda/2$, where $\lambda=c/f_\mathrm{c}$ is the carrier wavelength.
We use $10^4$ particles. The particles for the initial \ac{mt} states are drawn from a 5-D uniform distribution with center $\V{x}_{i,0} = [\V{p}_{i,0}^{\T}\;0\;\, 0;\,o_{i,0}]^{\T}\rmv$, where $\V{p}_{i,0}$ is the starting position of the actual \ac{mt} track, and the support of each position component about the respective center is given by $[-0.1\,\mathrm{m}, 0.1\,\mathrm{m}]$ and for each velocity component by $[-0.01\,\mathrm{m/s}, 0.01\,\mathrm{m/s}]$. If an IMU is used, the orientation state is represented by quaternions and are initialized using a 4-D Gaussian distribution with the center being the initial IMU measurements transformed to the quaternion representation and a standard deviation of  $10^\circ$ \cite{madgwick2010efficient}.
At time $n \rmv=\rmv 0$, the number of \acp{va} is $0$, i.e., no prior map information is available. 
The prior distribution for new \ac{pva} states $f_\text{n}\big(\overline{\V{x}}^{(j)}_{m,n}|\V{x}_n\big)$ is uniform on the square region given by $[-\text{45 m}, \text{45 m}] \times [-\text{45 m}, \text{45 m}]$ around the center of the floor plan shown in Fig.~\ref{fig:floorplan} and the mean number of new \acp{pva} at time $n$ is  $\mu_\text{n} = 0.01$. 
The probability of survival is $p_{\mathrm{s}} = 0.999$. The confirmation threshold as well as the pruning threshold are given as $p_{\mathrm{cf}} = 0.5$ and $p_{\mathrm{pr}} = 10^{-3}$, respectively. For the sake of numerical stability, we introduce a small amount of regularization noise to the \ac{va} state $\V{p}_{k,\mathrm{va}}$ at each time step $n$, i.e., $\underline{\V{p}}^{(j)}_{k,\mathrm{va}} \rmv\rmv=\rmv\rmv \V{p}^{(j)}_{k,\mathrm{va}} \rmv+\rmv \V{\omega}_{k}$, where $\V{\omega}_{k}$ is \ac{iid} across $k$, zero-mean, and Gaussian with covariance matrix $\sigma_a^2\, \V{I}_2$ and $\sigma_a = 10^{-3}\,\text{m}$. 
The \acp{mt} state transition for position and velocity is modeled by a constant-velocity and stochastic-acceleration model. In the case of IMU measurements, the stochastic acceleration term is replaced by the control term $\V{c}_{i,n-1}$\cite{madgwick2010efficient,HolDijLuiSch:ICUW2009}.
The state transition variances are set as $\sigma_\text{w} = 10^{-3}\,\mathrm{m/s^2}$.  The performance is measured in terms of the \ac{rmse} of the \ac{mt} position as well as the \acf{ospa} error \cite{Schuhmacher2008} of all \acp{va} with cutoff parameter and order set to 1~m and 2, respectively. 
The \ac{mospa} errors and \acp{rmse} of each unknown variable are obtained by averaging over all simulation runs. 

\subsubsection*{Experiment}
\label{sec:ResultsModel}
For the investigation of the proposed method, we define different experiments as shown in Table~\ref{tb:settings} and perform $100$ simulation runs for each setting.
\begin{table}[!t]
	\caption{Boolean table for experimental settings}
	\begin{center}
		\begin{tabular}{ c| l| l| l| l| l| l| l}
			&$E_1$ & $E_2$& $E_3$ & $E_4$ & $E_5$ &$E_6$ & $E_7$ \\
			\hline \hline
			 MIMO   & 0 & 1 & 1 & 0 & 0 & 1 & 1 \\ \hline
			 Coop   & 0 & 0 & 1 & 0 & 1 & 0 & 1 \\ \hline
			 IMU    & 1 & 1 & 0 & 1 & 1 & 1 & 1 \\ \hline
			 PVA Fusion & 0 & 0 & 1 & 1 & 1 & 1 & 1 \\ \hline
 \hline
		\end{tabular}
	\end{center}
	\label{tb:settings}
	\vspace*{-7mm}
\end{table} 
The results are summarized in Fig.~\ref{fig:results} and Fig.~\ref{fig:cdf}. In Fig.~\ref{fig:results}, we present the MOSPA as well as the RMSE of the estimated VAs and the cardinality error for four settings. The best result can be achieved using the proposed algorithm with PVA data fusion, cooperation, IMU measurements and MIMO system ($E_7$). A comparable result can be reached using the same settings but SIMO instead of MIMO ($E_5$). Using the proposed algorithm but without IMU measurements leads to a significant performance loss. This can be explained by the complex trajectory of MT $i=3$, since it can not be well described by the constant velocity motion model without an additional control term ($E_3$). Using the state-of-the-art \ac{Mpslam} with IMU measurements and the MIMO system but without PVA data fusion and cooperation ($E_2$) leads to the overall worst result, highlighting the significance of PVA fusion for robust and accurate mapping. 
Fig.~\ref{fig:cdf} shows the \ac{cf} of all MTs and all time steps for different experiments. The benefit of PVA data fusion and cooperative localization vs non-cooperative localization is clearly demonstrated for MIMO and SIMO systems ($E_6$ vs $E_7$ and $E_4$ vs $E_5$). As a comparison to the proposed method, we show the impact of not using PVA fusion on the MTs positioning error for SIMO and MIMO ($E_1$ and $E_2$). Interestingly the proposed method without IMU measurements ($E_3$) has the most outliers. This can be explained in since one of the MTs has a very complex trajectory, which often diverges due to missing navigation input. These errors are propagated through the cooperation and the PVA data fusion resulting in an overall diminished performance.

%% file: inputFiles/conclusions.tex
In this work, we present a novel approach for \ac{Mpslam} that performs data fusion over multiple observations of VAs by multiple MTs for robust and accurate mapping. This exchange of information for MTs also allows for cooperative localization in addition to multipath-based localization. We demonstrate, based on numerical simulations, the advantage of cooperative map fusion over state-of-the-art \ac{Mpslam} without map-fusion. In addition, we present the advantage of using additional IMU measurements to deal with complex trajectories.

%% file: inputFiles/Appendix.tex
\section{Signal Model} \label{app:signal}

Each \ac{bs} transmits at a center frequency $f_\text{c}$ an \ac{rf} signal $s^{(j)}_\text{bs}(t)$ with bandwidth $B$ and the \acp{mt} act as receivers. The \ac{rf} signal arrives at the receiver via the \ac{los} path as well as via \acp{mpc} originating from the reflections at surrounding objects. We assume time synchronization between all \acp{bs} and the \acp{mt}. However, our algorithm can be straightforwardly extended to an unsynchronized system according to \cite{LeitingerJSAC2015, GentnerTWC2016, EtzMeyHlaSprWym:TSP2017}.
We assume the \acp{bs} to use orthogonal codes, i.e., there is no mutual interference between individual \acp{bs}. Note that the proposed algorithm can be easily reformulated for the case where the \acp{mt} transmits \ac{rf} signals and the \acp{bs} act as receivers. The received signals at the $I$ \acp{mt} are synchronously sampled with sampling frequency given by the signal bandwidth $B$. In frequency domain, this results in $M = B/\Delta$ samples with frequency spacing $\Delta$. The discrete-frequency \ac{rf} signal model between \ac{bs} $j$ and \ac{mt} $i'$ is given by \cite{GreLeiWitFle:TWC2024}
\begin{align}\label{eq:signal_modelBSMT}
	{r_\text{bs}}_{n}^{(j,i')}[\ell,h,h'] &= \hspace{-3mm} \sum_{l=1}^{{N_\text{bs}}_n^{(j,i')}} \hspace{-2mm} \exp{\big(i 2 \pi \frac{f_\mathrm{c}}{c} d^{(j,h)} 
	\cos\big({\vartheta}^{(j,i')}_{l,n} -\psi^{(j,h)}\big)\big)} \nn \\
	& \hspace{-12mm}\times\exp{\big(i 2 \pi \frac{f_\mathrm{c}}{c} d^{(i',h')} 
	\cos\big({\theta}^{(j,i')}_{l,n}- o_{i',n} -\psi^{(i',h')}\big)\big)} \nn \\
	&  \hspace{-12mm} \times \V{s}({\tau_\text{bs}}_{l,n}^{(j,i')})[\ell] + \ist {w_\text{bs}}_n^{(j,i')}[\ell,h,h']
\end{align}
where  $\V{s}(\tau)[\ell] = {\alpha_\text{bs}}_{l,n}^{(j,i)} S^{(j)}_\text{bs}(\ell \Delta) \exp{\big(i 2\pi \ell \Delta \tau \big)} $, ${{N_\text{bs}}_n^{(j,i)}}$ is the number of visible path (related to \acp{va}), $S^{(j)}_\text{bs}(f)$ represents the signal spectrum (we assume here that $S^{(j)}_\text{bs}(f) = S_\text{bs}(f)$), $\ell =-(M-1)/2,\iist\ldots\iist,(M-1)/2$, $h$ represent the index for \ac{bs} antennas, $h'$ represents the index for \ac{mt} antennas. Delay, \ac{aoa} and \ac{aod} of the \ac{mpc} are represented by ${\tau_\text{bs}}_{l,n}^{(j,i)}$, $\theta^{(j,i)}_{l,n}$and $\vartheta^{(j,i)}_{l,n}$ respectively (see Fig.~\ref{fig:scenario}). The associated complex amplitude $\alpha_{l,n}^{(j,i)} \in \mathbb{C}$ is given by ${\alpha_\text{bs}}_{l,n}^{(j,i)} = {a_\text{bs}}^{(j,i)}_{l,n}\ist \text{e}^{-j2\pi f_\text{c}  {\tau_\text{bs}}_{l,n}^{(j,i)}} \frac{c}{4\pi f_\text{c} {d_\text{bs}}_{l,n}^{(j,i)} }$ where ${a_\text{bs}}^{(j,i)}_{l,n} \in \mathbb{C}$ is a reflection coefficient originating from all interactions of the radio signal with the radio equipment and the associated flat surfaces. The last term ${w_\text{bs}}_n^{(j,i)}[\ell,h]$ aggregates samples of the measurement noise and of a \ac{dmc} that incorporates the contributions from all other interactions, e.g. diffuse scattering and diffraction. It also includes low-power \acp{mpc} that cannot be resolved with the finite bandwidth \cite{GreLeiWitFle:TWC2024}.
The receive signal samples for \ac{bs}-\ac{mt} links are collected into the vector $\V{r}^{(j,i)}_{n}\in \mathbb{C}^{MHH'}$ as
\begin{align}\label{eq:sampledVector}
	\V{r}^{(j,i)}_{\text{bs}_n}  = & \big[\V{r}_{\text{bs}_n}^{(j,i)\ist\text{T}}[1,1] \hspace{1mm} \cdots \hspace{1mm} \V{r}_{\text{bs}_n}^{(j,i)\ist\text{T}}[H,1] \iist \V{r}_{\text{bs}_n}^{(j,i)\ist\text{T}}[1,2] \hspace{1mm} \cdots \hspace{1mm} \nn\\
	& \V{r}_{\text{bs}_n}^{(j,i)\ist\text{T}}[H,H']\big]^\T 
\end{align}
The discrete-frequency \ac{rf} signal model exchanged between \acp{mt} is given as $\V{r}^{(i,i')}_{\text{mt}_n}$ and can be obtained similarly to \eqref{eq:signal_modelBSMT} with the difference that the orientation of each MT has to be considered. The receive signal samples for \ac{mt}-\ac{mt} links are collected in the same manner as \eqref{eq:sampledVector}. We assume that the signal spectrum for MT-MT links is $S^{(i)}_{\text{mt}}(f) =S_{\text{bs}}(f)$.

\section{Data Association}\label{app:dataassoc}

\subsection{Base Stations}\label{sec:jointda}
At each time $n$ and for each \ac{bs} $j$, estimation of multiple \ac{pva} states is complicated by the data association uncertainty, i.e., it is unknown which measurement ${\RV{z}_\text{bs}}^{(j,i)}_{n}$ originated from which \ac{pva}. Furthermore, it is not known if a measurement did not originate from a \ac{pva} (i.e., a false alarm described by $ \mu_{\mathrm{fa}} $ and \ac{pdf} $ f_{\mathrm{fa}}({\V{z}_\text{bs}}^{(j,i)}_{m,n}) $), or if a \ac{pva} did not give rise to any measurement  missed detection). Following \cite{WilLau:J14,MeyKroWilLauHlaBraWin:J18,LeiMeyHlaWitTufWin:J19}, we assume that at any time $n$, each \ac{pva} can generate at most one measurement, and each measurement can be generated by at most one \ac{pva}. 
The associations between measurements and legacy \acp{pva} are described by the \ac{pva}-oriented association vector $ \underline{\RV{a}}_{n}^{(j,i)} \triangleq [\underline{\rv{a}}_{1,n}^{(j,i)} \ist \cdots \ist  \underline{\rv{a}}_{\rv{K}^{(j)}_{n-1},n}^{(j,i)}]^{\mathrm{T}} $ with entries 
%
$\underline{\rv{a}}_{k,n}^{(j,i)} \triangleq m \rmv\in\rmv {\Set{M}_\text{bs}}_n^{(j,i)}$ if legacy \ac{pva} $k$ generates measurement $m$ or $\underline{\rv{a}}_{k,n}^{(j,i)} \triangleq 0$ if legacy \ac{pva} $k$ does not generate any measurement.
In line with \cite{WilLau:J14,MeyKroWilLauHlaBraWin:J18,LeiMeyHlaWitTufWin:J19}, the associations can be equivalently described by a measurement-oriented association vector $ \overline{\RV{a}}_{n}^{(j,i)} \triangleq [\overline{\rv{a}}_{1,n}^{(j,i)} \ist \cdots \ist \overline{\rv{a}}_{{\rv{M}_\text{bs}}_n^{(j,i)},n}^{(j,i)}]^{\mathrm{T}} $ with entries 
%
${\overline{\rv{a}}}^{(j,i)}_{m,n} \triangleq k \rmv\in\rmv {\Set{K}}_n^{(j,i)}$ if measurement $m$ is originated by legacy \ac{pva} $k$ or ${\overline{\rv{a}}}^{(j,i)}_{m,n} \triangleq 0$ if measurement $m$ is not generated by any legacy \ac{pva} $k$.
The point target assumption is enforced by the exclusion function
%
$\Psi(\underline{\V{a}}_n^{(j,i)},\overline{\V{a}}_n^{(j,i)}) = \prod_{k = 1}^{K_{n-1}^{(j)}}\prod_{m = 1}^{{M_\text{bs}}_n^{(j,i)}}\psi(\underline{a}_{k,n}^{(j,i)},\overline{a}_{m,n}^{(j,i)})$
with $ \psi(\underline{a}_{k,n}^{(j)},\overline{a}_{m,n}^{(j)}) = 0 $, if $ \underline{a}_{k,n}^{(j)} = m $ and $ \overline{a}_{m,n}^{(j)} \neq k $ or $ \overline{a}_{m,n}^{(j)} = k $ and $ \underline{a}_{k,n}^{(j)} \neq m $, otherwise it equals $ 1 $. The ``redundant formulation'' of using $ \underline{\RV{a}}_{n}^{(j,i)} $ together with $ \overline{\RV{a}}_{n}^{(j,i)} $ is the key to making the algorithm scalable for large numbers of \acp{pva} and measurements \cite{MeyKroWilLauHlaBraWin:J18}. 
The vectors containing all association variables up to time $n$ are given for legacy \acp{pva} by $\underline{\RV{a}}_{1:n} \triangleq [\underline{\RV{a}}_{1}^{\mathrm{T}} \, ... \, \underline{\RV{a}}_{n}^{\mathrm{T}} ]^{\mathrm{T}}$ with $\underline{\RV{a}}_{n} \triangleq [\underline{\RV{a}}_{n}^{(1)\ist\mathrm{T}}\, ... \, \underline{\RV{a}}_{n}^{(J)\mathrm{T}} ]^{\mathrm{T}}$ and $\underline{\RV{a}}^{(j)}_{n} \triangleq [\underline{\RV{a}}_{n}^{(j,1)\ist\mathrm{T}}\, ... \, \underline{\RV{a}}_{n}^{(j,I)\mathrm{T}} ]^{\mathrm{T}}$ and for new \acp{pva} by $\overline{\RV{a}}_{1:n} \triangleq [\overline{\RV{a}}_{1}^{\mathrm{T}} \, ... \, \overline{\RV{a}}_{n}^{\mathrm{T}} ]^{\mathrm{T}}$ with $\overline{\RV{a}}_{n} \triangleq [\overline{\RV{a}}_{n}^{(1)\ist\mathrm{T}}\, ... \, \overline{\RV{a}}_{n}^{(J)\mathrm{T}} ]^{\mathrm{T}}$ and $\overline{\RV{a}}_{n}^{(j)}\triangleq [\overline{\RV{a}}_{n}^{(j,1)\ist\mathrm{T}}\, ... \, \overline{\RV{a}}_{n}^{(j,I)\mathrm{T}} ]^{\mathrm{T}}$.

\subsection{Cooperation} \label{sec:pda}
At each time $n$ and for each \ac{mt}-pair $(i,i')$ with $i \neq i'$, the measurements, i.e., the components of ${\RV{z}_\text{co}}^{(i,i')}_{m,n} \in {\mathcal{M}_\text{co}}^{(i,i')}_{n}$ are subject to the same data association uncertainty as mentioned before regarding \ac{los} identification and association. The association variable $\rv{b}^{(i,i')}_{n}$ for $i \neq i'$ is given by
%
$\rv{b}^{(i,i')}_{n} = m \in {\mathcal{M}_\text{co}}^{(i,i')}_{n}$ if LOS path is generated by measurement $m$ or $\rv{b}^{(i,i')}_{n} = 0$ otherwise.
We also define $\RV{b}_{1:n} \triangleq [\RV{b}_{1}^{\mathrm{T}} \, ... \, \RV{b}_{n}^{\mathrm{T}} ]^{\mathrm{T}}$ with $\RV{b}_{n} \triangleq [\rv{b}^{(1,2)\ist\mathrm{T}}_n \iist \rv{b}^{(1,3)\ist\mathrm{T}}_n \, \cdots \, \rv{b}^{(I,I)\ist\mathrm{T}}_n ]^{\mathrm{T}}$.